\documentclass[proceedings]{JHEP37}
\usepackage{eurosym}
\usepackage{amssymb}
\usepackage{amsfonts,cite}
\usepackage{amsmath}
\usepackage{graphicx}
\usepackage{multirow}
\usepackage{epsfig}

\newbox\mybox

\newcommand\fverb{\setbox\mybox=\hbox\bgroup\verb}
\newcommand\fverbdo{\egroup\medskip\noindent\fbox{\unhbox\mybox}\ }
\newcommand\fverbit{\egroup\item[\fbox{\unhbox\mybox}]}
\conference{Time-dependent C-operators as LR invariants in non-Hermitian theories}
\abstract{${\cal C}$-operators were introduced as involution operators in non-Hermitian theories that commute with the time-independent Hamiltonians and the parity/time-reversal operator. Here we propose a definition for time-dependent ${\cal C}(t)$-operators and demonstrate that for a particular signature they may be expanded in terms of time-dependent biorthonormal left and right eigenvectors of Lewis-Riesenfeld invariants. The vanishing commutation relation between the ${\cal C}$-operator and the Hamiltonian in the time-independent case is replaced by the Lewis-Riesenfeld equation in the time-dependent scenario. Thus, ${\cal C}(t)$-operators are always Lewis-Riesenfeld invariants, whereas the inverse is only true in certain circumstances. We demonstrate the working of the generalities for a non-Hermitian two-level matrix Hamiltonian. We show that solutions for ${\cal C}(t)$ and the time-dependent metric operator may be found that hold in all three ${\cal PT}$-regimes, i.e., the ${\cal PT}$-regime, the spontaneously broken ${\cal PT}$-regime and at the exceptional point. }

\title{Time-dependent C-operators as Lewis-Riesenfeld invariants in non-Hermitian theories}
\author{Andreas Fring, Takanobu Taira and Rebecca Tenney\\
 Department of Mathematics, City, University of London, Northampton Square,\\ London EC1V 0HB, UK \\
a.fring@city.ac.uk, takanobu.taira@city.ac.uk, rebecca.tenney@city.ac.uk}

\input{tcilatex}
\begin{document}

\section{Introduction}
By definition, quasi-Hermitian Hamiltonian systems are characterised by the intertwining relation $H^\dagger {\cal O} = {\cal O} H $ satisfied by their non-Hermitian Hamiltonian H \cite{Dieuqh,Urubu,Alirev}.  The operator ${\cal O}$ could be the indefinite parity operator ${\cal P}$ or the positive definite metric operator $\rho$. The ${\cal C}$-operator is then defined \cite{Bender:2002vv} as the multiplicative factor that converts the indefinite operator into a definite one, ${\cal P C} = \rho $. Time-independent ${\cal C}$-operators have been constructed and utilized in obtaining well-defined positive definite metric operators in many non-Hermitian theories, such as for instance theories with complex cubic interaction terms \cite{Bender:2004sa}, spin chain lattice models \cite{chainOla}, quantum field theories \cite{bender2004scalar,bender2005dual,fring2020goldstone}, non-Hermitian versions of quantum electrodynamics \cite{bender2005pt}, in semi-classical approximations \cite{bender2004semi} and in non-Hermitian theories modelling superconductivity with ${\cal PT}$-symmetric Cooper pairing symmetry \cite{ghatak2018theory}. In principle, many more models for which the metric operator $\rho$ has been constructed by directly solving the quasi-Hermiticity relation could be listed here as the parity operator ${\cal P } $ can usually be identified trivially so that one immediately obtains the ${\cal C}$-operators from ${\cal C} = {\cal P}\rho $. For instance in \cite{JM,ACIso,Mostsyme,PEGAAF,MGH,PEGAAF2} exact solutions for the metric in systems with underlying infinite dimensional Hilbert spaces have been found and the corresponding ${\cal P}$-operators were identified. We stress here that the ${\cal C}$-operator should not be confused with the charge conjugation operator that maps particles to anti-particles and vice versa in quantum field theoretical systems, even though some of their general properties are shared.

While the scheme of pseudo/quasi-Hermitian ${\cal PT}$-symmetric systems \cite{Bender:1998ke,Alirev,bagarello2015non,PTbook} has been generalised to explicitly time-dependent Hamiltonian systems 
 \cite{CA,CArev,time1,time6,BilaAd,fringmoussa,fringmoussa2,time7,mostafazadeh2020time,AndTom1,AndTom2,AndTom3,AndTom4,AndTom5,cen2019time,fring2019eternal,fring2020time,frith2020exotic,BeckyAnd2,fring2021perturb,fring2021exactly,fring2021infinit}, time-dependent versions of ${\cal C}$-operators have not been discussed so far. We introduce here a definition for time-dependent ${\cal C}$-operators and study their properties. We find that the factorisation property ${\cal C}(t) = {\cal P} \rho(t) $, its involutory nature, ${\cal C}^2(t) = \mathbb{I}$, and the vanishing commutation relation with the ${\cal PT}$-operator are preserved in the time-dependent setting. However, crucially we will argue here that time-dependent ${\cal C}(t)$-operators no longer commute with the Hamiltonian. We will demonstrate that instead they have to satisfy the Lewis-Riesenfeld equation \cite{Lewis69}. This means that time-dependent ${\cal C}(t)$-operators are in fact Lewis-Riesenfeld invariants. The latter have been devised originally to facilitate the construction of solutions to the time-dependent Schr\"odinger equation, as they reduce the former to an eigenvalue problem. For time-dependent non-Hermitian systems they were also found to be extremely useful, since they can be utilised to reduce the time-dependent Dyson equation, that is a first order differential equation, to as a much simpler similarity relation \cite{maamache2017pseudo,khantoul2017invariant,AndTom4,cen2019time,frith2019time,fring2021exactly}. 
 
 Our manuscript is organised as follows: In section 2 we first recall the key features of time-independent versions of ${\cal C}$-operators, propose a definition for their time-dependent versions ${\cal C}(t)$ and study their general properties in particular we demonstrate how they are related to Lewis-Riesenfeld invariants. In section 3 we illustrate the working of the general formulae for a two level system, first in a time-independent setting, a scenario with time-independent Hamiltonian, but time-dependent metric and finally in a fully fledged time-dependent scenario. Our conclusions are stated in section 4.

\section{${\cal C}$-operators and invariants}
\subsection{Time-independent ${\cal C}$-operators}

Before proposing a definition for time-dependent ${\cal C}$-operators we recall their definition in the time-independent case and briefly discuss the role they play. When dealing with non-Hermitian Hamiltonians, $H \neq H^\dagger$, with a discrete spectrum standard orthonormal basis have to be replaced with biorthonormal basis comprised of their left and right eigenvectors, $ \vert \Phi \rangle$ and $\vert \Psi \rangle$, respectively. They are defined by the right and left eigenvalue equations
\begin{equation}
	H \vert \Psi_n \rangle = E_n \vert \Psi_n \rangle , \qquad H^\dagger \vert \Phi_n \rangle = E_n \vert \Phi_n \rangle, \qquad n \in \mathbb{N} ,  \label{leftright} 
\end{equation}
respectively. These eigenvectors are biorthonormal to each other and are complete \cite{dieudonne1953bio,Weigertbi} 
\begin{equation}
	\langle \Phi_n \vert \Psi_m \rangle =  \langle \Psi_n \vert \Phi_m \rangle = \delta_{nm}, \qquad \sum_n \vert \Phi_n \rangle \langle \Psi_n \vert = \sum_n \vert \Psi_n \rangle \langle \Phi_n \vert =\mathbb{I} . \label{36}
\end{equation}  
The biorthonormal basis can be employed to introduce a new so-called ${\cal C}$-operator  \cite{Bender:2002vv}
\begin{equation}
	{\cal{C}} :=  \sum_n s_n  \vert \Psi_n \rangle \langle \Phi_n \vert,  \qquad s_n= \pm 1, \label{defC}
\end{equation}
with the set $\{s_1, \ldots, s_n\}$ being the signature. In turn, this operator equals the product of the indefinite parity operator ${\cal P}$ and the positive-definite Hermitian metric $\rho$  
\begin{equation}
  {\cal C}	= {\cal P} \rho ,     \label{rPC} 
\end{equation}
with properties
 \begin{equation}
 	H^\dagger \rho	=\rho H  ,  \quad \rho^\dagger =\rho \qquad \text{and}  \qquad   H^\dagger {\cal P}	= {\cal P} H  , \quad {\cal{P}}^2  =\mathbb{I}. \label{rpp} 
 \end{equation}
The Hermiticity of $\rho (t)$ implies that ${\cal C}$ is pseudo-Hermitian with regard to the adjoint action of the parity operator ${\cal P}$. This follows by conjugating (\ref{rPC}) 
\begin{equation}
	{\cal C}^\dagger= \rho^\dagger {\cal P}=  {\cal P} {\cal P} \rho {\cal P} = 
	{\cal P} {\cal C} {\cal P} .  \label{pseudoC}
\end{equation}
As a consequence of the previous relations one obtains the three constraints
\begin{equation}
	{\cal C}^2= \mathbb{I}, \qquad \left[ {\cal PT}, {\cal C}   \right]=0, \qquad \left[ H, {\cal C}   \right] =0 . \label{algC}
\end{equation}
The first property simply follows from the defining relation (\ref{defC}) together with the orthonormality relation for the left and right eigenvectors (\ref{36}). The second equation in (\ref{algC}) follows by multiplying (\ref{pseudoC}) from the right by the time-reversal operator ${\cal T}$ from the right and using the fact that ${\cal C}^\dagger {\cal T}= {\cal T} {\cal C} $. The third relation follows by multiplying the first equation in (\ref{rpp}) from the right by ${\cal P} $  together with the last two relations in (\ref{rpp}) and (\ref{rPC}).
\begin{equation}
	{\cal P} H^\dagger \rho	= {\cal P} \rho H \quad \Rightarrow \quad {\cal P} H^\dagger {\cal PC }	= {\cal P} {\cal PC } H \quad \Rightarrow \quad H {\cal C } = {\cal C }H. 
\end{equation}
Instead of employing the left and right eigenvectors to obtain an explicit expression for the ${\cal C }$-operator, it was suggested in \cite{Bender:2004sa} that one may also solve the equation (\ref{algC}) to construct it algebraically. Especially for systems with an infinite dimensional Hilbert space this is advantageous, as in the definition (\ref{defC}) one is even in the rare cases of exactly solvable models left with an infinite sum. See \cite{Bender:2004sa,chainOla,bender2004scalar,bender2005dual,fring2020goldstone,bender2005pt,bender2004semi,ghatak2018theory} for examples where this algebraic approach has been carried out in one form or another. 

While $\rho$ and ${\cal P}$ satisfy the same equations, the metric operator is positive definite, whereas the parity operator is in general not positive definite. Thus we may also read equation (\ref{rPC}) as $  {\cal P}{\cal C}	= \rho$, so that the ${\cal C}$-operator can be interpreted as the operator that by multiplication converts the indefinite operator ${\cal P}$ into a positive definite operator $\rho$, which in turn serves to define the $\rho$-inner product
$ \left\langle  \Psi \right\vert \tilde{\Psi} \rangle_\rho :=  \left\langle  \Psi \right\vert  \rho \tilde{\Psi} \rangle  $. Moreover, the metric can be decomposed as $\rho=\eta^\dagger \eta$ where $\eta$ is the Dyson map that adjointly maps the non-Hermitian Hamiltonian $H$ to a Hermitian one $h=h^\dagger= \eta H \eta^{-1}$, by relating the eigenstates $ \eta \vert \Psi \rangle = \vert \phi \rangle$ of the corresponding eigenvalue equation $h  \vert \phi \rangle = E  \vert \phi \rangle  $.

\subsection{Time-dependent ${\cal C}$-operators}
Let us now see how the above is generalised and the ${\cal C}$-operator is naturally introduced for the time-dependent setting. For explicitly time-dependent systems we propose the definition
\begin{equation}
	{\cal{C}}(t) :=  \sum_n s_n  \vert \Psi_n(t) \rangle \langle \Phi_n(t) \vert,  \qquad s_n= \pm 1, \label{defCt}
\end{equation}
and assume that the states $ \vert \Psi_n(t) \rangle$ and  $\vert \Phi_n(t) \rangle$ satisfy the time-dependent Schr\"odinger equations
\begin{equation}
	i \hbar \partial_t \vert \Psi_n(t) \rangle = H(t) \vert \Psi_n(t) \rangle , \qquad \text{and} \qquad i \hbar \partial_t \vert \Phi_n(t) \rangle = H^\dagger(t) \vert \Phi_n(t) \rangle . \label{TDSE}
\end{equation}
Below we discuss how the discretisation of these states is to be understood. The first property in (\ref{algC}) follows now by the same reasoning as in the time-independent case by using the defining relation (\ref{defCt}) and assuming that (\ref{pseudoC}) still holds for the time-dependent case, also the second relation follows. However, the third constraint needs modification as is seen by differentiating (\ref{defCt}) with respect to time 
\begin{equation}
	i \hbar \partial_t {\cal{C}}(t)  =  i \hbar \sum_n s_n \left\{ \partial_t \vert \Psi_n(t) \rangle \langle \Phi_n(t) \vert +   \vert \Psi_n(t) \rangle \partial_t \langle \Phi_n(t) \vert  \right\}
	=  [H,{\cal{C}}(t)] , \label{CLR}
\end{equation}
where we simply used the time-dependent Schr\"odinger equations (\ref{TDSE}). Thus ${\cal{C}}(t)$ is a conserved quantity, as we may read equation (\ref{CLR}) as $d {\cal{C}}(t)/dt =0 $, when invoking Heisenberg's equation of motion. To summarize, the three constraining relations (\ref{algC}) underlying the algebraic approach in the time-independent case have to be replaced by
\begin{equation}
	{\cal C}^2(t)= \mathbb{I}, \qquad \left[ {\cal PT}, {\cal C}(t)   \right]=0, \qquad \left[ H, {\cal C}(t)   \right] =i \hbar \partial_t {\cal{C}}(t), \label{algCt}
\end{equation}
in the time-dependent scenario.

Next we establish how the time-dependent ${\cal{C}}$-operator relates to the time-dependent metric and demonstrate that the relation (\ref{rPC}) directly generalises to the time-dependent scenario as
\begin{equation}
	{\cal C}(t)	= {\cal P} \rho(t) .     \label{rPCt} 
\end{equation}
This follows by solving (\ref{rPCt}) for the metric and differentiating the resulting equation with respect to $t$. In this way we compute
\begin{eqnarray}
	i\hbar \partial _{t}\rho (t) &=&  {\cal P} i\hbar \partial_t {\cal C}(t) \\
	&=& {\cal P} H {\cal C}(t) - {\cal P} {\cal C}(t) H \\
	&=&  H^\dagger {\cal P} {\cal C}(t) - {\cal P} {\cal C}(t) H \\
	&=& H^{\dagger }\rho (t)-\rho (t)H ,
\end{eqnarray}
which is precisely the time-dependent quasi-Hermiticity relation that generalises (\ref{rpp}), see \cite{CA,CArev,time1,time6,BilaAd,fringmoussa,fringmoussa2,time7,mostafazadeh2020time,AndTom1,AndTom2,AndTom3,AndTom4,AndTom5,cen2019time,fring2019eternal,fring2020time,frith2020exotic,BeckyAnd2,fring2021perturb,fring2021exactly,fring2021infinit}. Thus (\ref{rPCt}) holds consistently for the definition (\ref{defCt}) together with the properties of the parity operator as stated in (\ref{rpp}).

\subsection{Lewis-Riesenfeld invariants}
We notice that (\ref{CLR}) is identical to the defining relation for the Lewis-Riesenfeld invariants $I_H(t)$, the Lewis-Riesenfeld equation  
\begin{equation}
		i \hbar \partial_t I_H(t)  = [H,I_H(t)] , \label{LRdef}
\end{equation}	
suggesting therefore a possible relation between the two. In general, the main advantage of employing these invariants is that they reduce the time-dependent Schr\"odinger equation to an eigenvalue problem in which time simply plays the role of a standard parameter. The invariants satisfy the eigenvalue equations
\begin{equation}
	I_H(t)\left\vert \Psi ^I(t)\right\rangle
	= \Lambda \left\vert \Psi ^I(t)\right\rangle , \quad
	\left\vert \Psi (t)\right\rangle =  e^{i \hbar \alpha(t)}\left\vert \Psi ^I(t)\right\rangle, \quad \partial_t \Lambda =0, \label{LR2}
\end{equation}  
where the argument of the phase factor, that relates the solution of the time-dependent Schr\"odinger equation to the eigenstates of the Lewis-Riesenfeld invariants, can be determined from $  \dot{\alpha}=   \left\langle  \Psi ^I(t)    \right\vert   i  \partial_t - H(t)/\hbar     \left\vert \Psi ^I(t)\right\rangle_\rho$. 

Since for the non-Hermitian Hamiltonain $H$ the invariant $I_H(t)$ must also be non-Hermitian, it possesses in addition to the set of right eigenvectors in (\ref{LR2}) a set of left eigenvectors  $\left\vert \Phi ^I(t)\right\rangle$ with
\begin{equation}
	I_H^\dagger(t)\left\vert \Phi ^I(t)\right\rangle
	= \Lambda \left\vert \Phi ^I(t)\right\rangle , \quad
	\left\vert \Phi (t)\right\rangle =  e^{i \hbar \alpha(t)}\left\vert \Phi ^I(t)\right\rangle. \label{LR3}
\end{equation}  
We assume now that the invariant $I_H(t)$ possesses a discrete spectrum with eigenvalues $\Lambda_n$ and eigenfunctions $\left\vert \Psi_n ^I(t)\right\rangle$. This discretisation is then inherited by the solutions of the time-dependent Schr\"odinger equations in (\ref{TDSE}) via (\ref{LR2}) and
an immediate consequence of these relations is that we can also expand the time-dependent ${\cal{C}}$-operator in terms of the left and right eigenstates of the Lewis-Riesenfeld invariant 
\begin{equation}
	{\cal{C}}(t) :=  \sum_n s_n  \vert \Psi^I_n(t) \rangle \langle \Phi^I_n(t) \vert,  \qquad s_n= \pm 1. \label{defCt2}
\end{equation}
The phases cancel out in (\ref{defC}) so that the solutions of the time-dependent Schr\"odinger equation are replaced by the eigenstates of the Lewis-Riesenfeld invariants. This means all time-dependent ${\cal{C}}$-operators are also Lewis-Riesenfeld invariants. The inverse does not always hold and one may easily construct invariants that are not ${\cal{C}}$-operators. However, noting that we may expand the invariants as  
\begin{equation}
	I_h(t) :=  \sum_n \Lambda_n  \vert \Psi^I_n(t) \rangle \langle \Phi^I_n(t) \vert,  
\end{equation}
we can obviously achieve the equality ${\cal{C}}(t)=I_H(t)$, if we can tune the eigenvalues for the invariant such that $\{ \Lambda_1, \ldots , \Lambda_n \}= \{s_1, \ldots , s_n \}$. This may be achieved by enforcing the first two relations in (\ref{algCt}).

\section{A two level system}
In order to illustrate the consistent working of the proposed formulae above we will present a simple worked out example i) in the time-independent case, ii) for time-independent Hamiltonian and time-dependent metric and iii) for the fully time-dependent scenario. 

As many techniques have been devised to construct Lewis-Riesenfeld invariants \cite{maamache2017pseudo,khantoul2017invariant,AndTom4,cen2019time,frith2019time,fring2021exactly},
their construction will be our starting point in finding time-dependent ${\cal{C}}(t)$-operators from which we subsequently compute the metric operators by means of (\ref{rPC}). We briefly explain one general method by considering the most general two level matrix Hamiltonian and invariant expanded in terms of Pauli matrices $\sigma_{x,y,z}$ as 
\begin{equation}
	H(t)= h_0(t) \mathbb{I} +  h_1(t) \sigma_x + h_2(t) \sigma_y +h_3(t) \sigma_z, \quad  
	I_H(t)= \iota_0(t) \mathbb{I} +  \iota_1(t) \sigma_x + \iota_2(t) \sigma_y +\iota_3(t) \sigma_z, \label{hinv}
\end{equation}
with time-dependent coefficient functions $h_i(t),\iota_i(t) \in \mathbb{C} $, $i=0,1,2,3$. When substituting these expressions into the Lewis-Riesenfeld equation (\ref{LRdef}), it reduces to
\begin{equation}
   \partial_t \iota_0=0, \qquad	\partial_t \vec{\iota}  = \frac{2}{\hbar} M \vec{\iota},\quad \text{with} \quad M_{ij} = - \epsilon_{ijk} h_k = \left(
	\begin{array}{ccc}
		0 & -h_3 & h_2 \\
		h_3 & 0 & -h_1 \\
		-h_2 & h_1 & 0 \\
	\end{array}
	\right)_{ij} . \label{matequ}
\end{equation}
The general solution of (\ref{matequ}) is then
\begin{eqnarray}
 \vec{\iota}(t)  &=& T \exp\left[ \frac{2}{\hbar} \int_0^t M(s) ds  \right] \vec{\iota}(0)  \label{gensolM} \\
 &=&  \sum_{n=0}^{\infty} \left( \frac{2}{\hbar}  \right)^n T  \left[ \int_0^t M(t_1) dt_1 \int_0^{t_1} M(t_2) dt_2 \ldots \int_0^{t_{n-1}} M(t_{n-1}) dt_n \right] \vec{\iota}(0) , \label{timeord}
\end{eqnarray}
with $t> t_1> \ldots t_n >0$ and $T$ denoting the time-ordering operator. In what follows we will consider some special cases of this solution for which the time-ordered exponential can be computed explicitly. 

\subsection{A time-independent system}
For reference purposes we start with a well-known two level example for which all of the above quantities can be calculated easily	
\begin{equation}
	H=-\frac{1}{2}\left( \omega \mathbb{I}+\lambda \sigma _{z}+i\kappa \sigma
	_{x}\right)=-\frac{1}{2}\left(
	\begin{array}{cc}
		\omega + \lambda   & i \kappa  \\
		i \kappa  & \omega -\lambda  \\
	\end{array}
	\right), \qquad \omega, \lambda, \kappa \in \mathbb{R}.   \label{Hfinite}
\end{equation}
The eigenenergies together with the left and right eigenvectors are obtained as
\begin{equation}
	E_{\pm }=-\frac{1}{2}\omega \pm \frac{1}{2}\sqrt{\lambda ^{2}-\kappa ^{2}}%
	,\quad  \vert \Psi _{\pm } \rangle= \frac{1}{\sqrt{N_{\pm}}}  \left( 
	\begin{array}{c}
		i\left(-\lambda \pm \sqrt{\lambda ^{2}-\kappa ^{2}} \right) \\ 
		\kappa%
	\end{array}%
	\right) , \quad \vert \Phi _{\pm } \rangle= \mp {\cal P} \vert \Psi _{\pm } \rangle.   \label{eigenv}
\end{equation}
The parity operator is identified from (\ref{rpp}) as ${\cal P}= \sigma_z $ and the normalisation constant $N_{\pm}= 2 (\lambda  \sqrt{\lambda ^2-\kappa ^2}\pm\kappa ^2\mp\lambda ^2)$ is determined by (\ref{36}). For $ \vert \lambda \vert > \vert \kappa \vert$ these states are ${\cal PT}$ symmetric, ${\cal PT} \vert \Psi _{\pm } \rangle=- \vert \Psi _{\pm } \rangle $ when we identify ${\cal PT}:= \sigma_z \tau$ with $\tau$ being a complex conjugation. The Hamiltonian respects the same symmetry, i.e., $\left[ H, {\cal PT} \right] =0$. 

With signature $\{+,-\}$, the ${\cal C}$-operator is directly computed from the defining relation (\ref{defC})  
\begin{equation}
	{\cal C} =\frac{1}{\sqrt{\lambda^2 -\kappa^2}}\left(
	\begin{array}{cc}
		\lambda  & i \kappa  \\
		i \kappa  & -\lambda  \\
	\end{array}
	\right).
\end{equation}
Thus the metric is $\rho = {\cal P C} $, from which one obtains the Dyson map $\eta$ by solving $\rho = \eta^\dagger \eta$. We can also obtain $\eta$ directly by defining it in term eigenvectors of $H$ as column vectors $\eta= \{\psi_+,\psi_-\}^\intercal$, since the adjoint action of this operator will always diagonalise the Hamiltonian with $h=\text{diag}\{E_+,E_-\}$.

\subsection{The metric picture}
When introducing a time-dependence, an interesting new option emerges that does not exist in the time-independent case, which allows to include the time-dependence into the metric $\rho(t)$, while keeping the Hamiltonian still time-independent. We now demonstrate that the ${\cal C}$-operator resulting in this case is in fact identical to the Lewis-Riesenfeld invariant. For this purpose we present a simple solution to the Lewis-Riesenfeld equation (\ref{CLR}) for the time-independent Hamiltonian (\ref{Hfinite}) resulting from the general scheme (\ref{hinv})-(\ref{timeord}). Setting in equation (\ref{hinv}) the time-dependent coefficient functions to constants $h_0(t) = -1/2 \omega$, $h_1(t) = -i/2 \kappa $, $h_2(t) =0$, $h_3(t) = -1/2 \lambda $, we obtain the time-independent Hamiltonian in (\ref{Hfinite}). Next we construct from (\ref{gensolM}) invariants that are all of the form
\begin{equation}
	I_H(t) =\frac{1}{\xi}\left(
	\begin{array}{cc}
		-\delta  & \gamma _+ \\
		\gamma _- & \delta  \\
	\end{array}
	\right), \label{invgen}
\end{equation}  
with $\det[	I_H(t)]=-1$ by making different choices for the initial conditions. Choosing $\iota_0(t)=0$ and $\vec{\iota}(0)= \{i \sqrt{2} \kappa /\xi , i, \sqrt{2} \lambda /\xi  \} $ the evaluation of (\ref{gensolM}) leads to an invariant (\ref{invgen}) in the ${\cal PT}$-symmetric regime with
\begin{equation}
	{\cal PT}: \xi = \sqrt{\lambda ^2-\kappa ^2}, \,\, \delta= -\sqrt{2}\lambda - \kappa  \sin (\xi  t), \,\, \gamma_{\pm}= \pm \xi \cos(\xi t) + i \left[\sqrt{2} \kappa +  \lambda  \sin (\xi  t) \right]. \label{IPT}
\end{equation}
Taking instead $\iota_0(t)=0$ and $\vec{\iota}(0)= \{ i (\sqrt{2}\lambda- \kappa) /\xi , 0, (\sqrt{2} \kappa -\lambda) /\xi  \} $ we obtain a solution for the spontaneously broken ${\cal PT}$-regime with
\begin{equation}
	b{\cal PT}: \xi = \sqrt{\kappa ^2-\lambda ^2}, \,\, \delta= \lambda -\sqrt{2} \kappa  \cosh (\xi  t), \,\, \gamma_{\pm}= \pm \sqrt{2} \xi  \sinh (\xi  t) + i \left[ \sqrt{2} \lambda  \cosh (\xi  t)- \kappa \right] . \label{IbPT}
\end{equation}
Setting $\lambda=\kappa$ with initial conditions $\iota_0(t)=0$ and $\vec{\iota}(0)= \{ 0,i,\sqrt{2}  \} $ the solution (\ref{gensolM}) becomes valid at the exceptional point with
\begin{equation}
\text{EP}: \xi = 1, \,\, \delta= -\frac{\kappa ^2 t^2}{\sqrt{2}}-\kappa  t -\sqrt{2}, \,\, \gamma_{\pm}=\pm\left( 1+\sqrt{2} \kappa  t \right) +i \left[\frac{\kappa ^2 t^2}{\sqrt{2}}+\kappa  t\right] . \label{IEP}
\end{equation}
The most general solution involves four integration constants corresponding to the initial conditions, which have been chosen here conveniently to keep our expressions simple. 

Since the invariant $I_H(t)$ is associated to a non-Hermitian Hamiltonian, it is by (\ref{LRdef}) itself also non-Hermitian, and therefore possesses a set of left and right eigenvectors that form a biorthonormal basis
\begin{equation}
  \vert \Psi _{\pm } \rangle= \frac{1}{\sqrt{ N_{\pm}}}  \left( 
	\begin{array}{c}
		\delta\mp\xi \\ 
	      - \gamma_-
	\end{array}%
	\right) , \quad \vert \Phi _{\pm } \rangle=  {\cal P} \vert \Psi _{\pm } \rangle ,  \quad I_H(t) \vert \Psi _{\pm } \rangle = \pm \vert \Psi _{\pm } \rangle,  \quad I^\dagger_H(t) \vert \Phi _{\pm } \rangle = \pm \vert \Phi _{\pm } \rangle,  \label{eigenvI}
\end{equation}
satisfying the biorthonormality relations (\ref{36}), with normalisation constants $N_{\pm} = 2 (\xi^2 \mp \delta \xi)$. Using the definition (\ref{defC}) of the time-dependent ${\cal C}$-operator, we obtain
\begin{equation}
	{\cal C}(t)=  \vert \Psi _{+} \rangle  \langle \Phi _{+} \vert - \vert \Psi _{-} \rangle  \langle \Phi _{-} \vert = I_H(t) . \label{CI}
\end{equation}
We verify that the pseudo-Hermitian relation (\ref{pseudoC}) holds for this operator. For the signatures $\{\pm,\pm\}$ we simply obtain the trivial solutions $	{\cal C}(t)=I_H(t)=\pm \mathbb{I}$ and evidently for $\{-,+\}$ equations (\ref{CI}) is just multiplied by $-1$. Thus, we have found ${\cal C}$-operators identical to the Lewis-Riesenfeld invariant in all three ${\cal PT}$-regimes. 

 From relation (\ref{rPCt}) we easily obtain a metric in each of the three regimes
\begin{eqnarray}
	\rho_{b{\cal PT}}(t) &=& \frac{1}{\xi} \left(
	\begin{array}{cc}
		\sqrt{2} \kappa  \cosh (\xi  t)-\lambda  & \sqrt{2} \xi  \sinh (\xi  t) -i \left[ \kappa + \sqrt{2} \lambda  \cosh (\xi  t) \right] \notag \\
		\sqrt{2} \xi  \sinh (\xi  t) +i \left[ \kappa + \sqrt{2} \lambda  \cosh (\xi  t) \right]  &  \sqrt{2} \kappa  \cosh (\xi  t) -\lambda \\
	\end{array}
	\right) , \\
	\rho_{{\cal PT}}(t) &=& \frac{1}{\xi} \left(
	\begin{array}{cc}
		\sqrt{2} \lambda +\kappa  \sin (\xi  t) & \xi  \cos (\xi  t)+i \left[\sqrt{2} \kappa +\lambda  \sin (\xi  t)\right] \\
		\xi  \cos (\xi  t)-i \left[\sqrt{2} \kappa +\lambda  \sin (\xi  t)\right] & \sqrt{2} \lambda +\kappa  \sin (\xi  t) \\
	\end{array}
	\right), \label{rho2} \\
	\rho_{\text{EP}}(t) &=& \left(
	\begin{array}{cc}
		\frac{\kappa ^2 t^2}{\sqrt{2}}+\kappa  t+\sqrt{2} & 1+\sqrt{2} \kappa  t+\frac{1}{2} i \kappa  t \left(\sqrt{2} \kappa  t+2\right) \\
		1+\sqrt{2} \kappa  t-\frac{1}{2} i \kappa  t \left(\sqrt{2} \kappa  t+2\right) & \frac{\kappa ^2 t^2}{\sqrt{2}}+\kappa  t+\sqrt{2} \\
	\end{array}
	\right), \notag
\end{eqnarray}
which are Hermitian as long as $\xi \in \mathbb{R}$, which is guaranteed for the appropriate choices of $\xi$ in the different regimes. Moreover, $\rho$ is positive definite in these regimes with $\det \rho(t) = 1$. Thus in any of the three ${\cal PT}$-regimes one may also find a well-defined metric from a given Lewis-Riesenfeld invariant/${\cal C}$-operator and parity operator. However, we notice that we can not cross over smoothly from one ${\cal PT}$-regime to the other, so that the three solutions in (\ref{rho2}) correspond to different theories. As we will see in the next section this is not always the case and one can also construct solutions for {\em one} theory that is defined in all three regimes. 

\subsection{Fully Time-dependent scenario}
By slightly modifying (\ref{Hfinite}), let us now consider the explicitly time-dependent Hamiltonian
\begin{equation}
	H=-\frac{1}{2}\left( \omega \mathbb{I}+\lambda \tau(t) \sigma _{z}+i\kappa \tau(t) \sigma
	_{x}\right)=-\frac{1}{2}\left(
	\begin{array}{cc}
		\omega + \lambda \tau(t)   & i \kappa \tau(t)  \\
		i \kappa \tau(t)  & \omega -\lambda \tau(t)  \\
	\end{array}
	\right), \qquad \omega, \lambda, \kappa, \tau(t) \in \mathbb{R}.  \label{Hfinitet}
\end{equation} 
Setting in equation (\ref{hinv}) the time-dependent coefficient functions to $h_0(t) = -1/2 \omega$, $h_1(t) = -i/2 \kappa \tau(t)$, $h_2(t) =0$, $h_3(t) = -1/2 \lambda \tau(t)$, $\iota_0(t)=0$ we obtain the time-dependent Hamiltonian in (\ref{Hfinitet}). Furthermore when choosing the initial condition in (\ref{gensolM})  to $\vec{\iota}(0)= \{0,0,1 \}$, we find a simple invariant $I_H(t)$ of the same general form as in (\ref{invgen}) with re-defined entries
\begin{equation}
	\xi = \kappa ^2-\lambda ^2, \,\, \delta =\lambda ^2-\kappa ^2 \cosh [\mu (t)], \,\, \gamma_\pm = \pm \kappa  \sqrt{\xi} \sinh [\mu (t)]+i \kappa  \lambda  \{\cosh [\mu (t)]-1\}, \label{dgamma}
\end{equation}
where $\mu(t)= \sqrt{\xi} \int^t \tau (s) \, ds$. Unlike the invariants (\ref{IPT}), (\ref{IbPT}) and (\ref{IEP}), this invariant is meaningful in all three ${\cal PT}$-regimes. Moreover, the expressions for the left and right eigenstates (\ref{eigenvI}) still hold with (\ref{dgamma}). Hence we have also in this case the relation (\ref{CI}) and the ${\cal C}$-operator is identical to the Lewis-Riesenfeld invariant. 

A Hermitian positive definite metric operator valid in all three different ${\cal PT}$-regimes is then obtained from (\ref{rPCt}) as 
\begin{equation}
	\rho(t)=\frac{1}{\xi} \left(
	\begin{array}{cc}
		\kappa ^2 \cosh [\mu (t)]-\lambda ^2 & \kappa  \sqrt{\xi } \sinh [\mu (t))]+i \kappa  \lambda  \{\cosh [\mu (t)]-1\} \\
		\kappa  \sqrt{\xi } \sinh [\mu (t)]-i \kappa  \lambda  \{\cosh [\mu (t)]-1\} & \kappa ^2 \cosh [\mu (t)]-\lambda ^2 \\
	\end{array}
	\right) . \label{rhof}
\end{equation}
The left and right limit to the exceptional point, that is approaching either from the ${\cal PT}$-symmetric or the spontaneously broken regime, can be carried out in a smooth manner
\begin{equation}
\lim_{\lambda \uparrow \kappa}	\rho(t)=\lim_{\lambda \downarrow \kappa}	\rho(t)=\left(
\begin{array}{cc}
	1+\frac{\kappa ^2 \tilde{\mu }^2}{2} &  \kappa  \tilde{\mu } +i \frac{\kappa ^2 \tilde{\mu }^2}{2} \\
	\kappa  \tilde{\mu } -i \frac{\kappa ^2 \tilde{\mu }^2}{2} & 1+\frac{\kappa ^2 \tilde{\mu }^2}{2} \\
\end{array}
\right) , \label{rhofff}
\end{equation}
where $\tilde{\mu }(t)= \int^t \tau(s) ds$.

Notice that these solutions also hold for $\tau(t)=1$, in which case we obtain a solution for $\rho(t)$ in the metric picture with time-independent Hamiltonian valid in all three ${\cal PT}$-regimes. 

\begin{figure}[h]
	\noindent	\begin{minipage}[b]{0.5\textwidth}     \!\!\!\! \!\!\!\! \includegraphics[width=\textwidth]{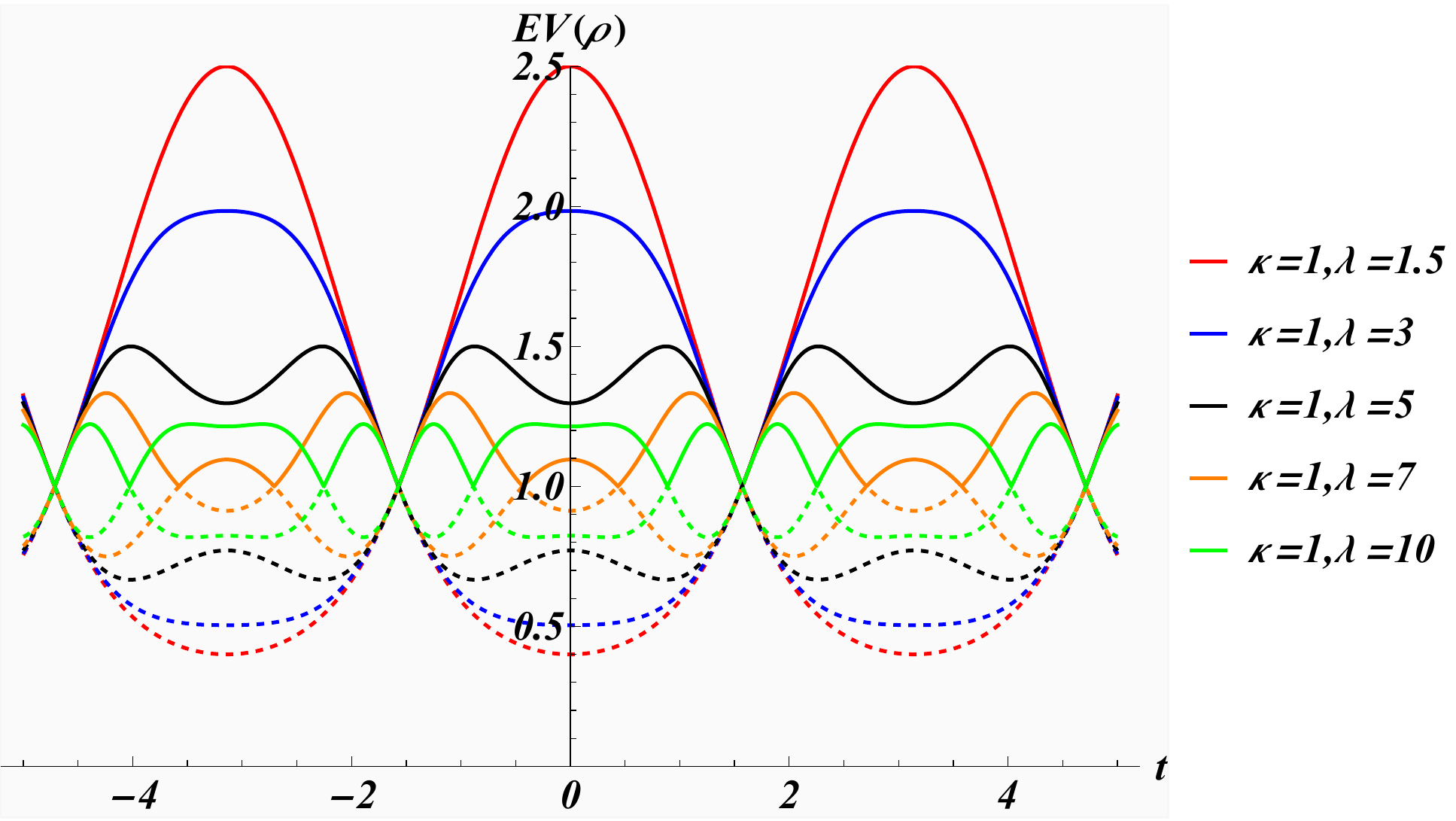}
	\end{minipage}
	\begin{minipage}[b]{0.5\textwidth}      \includegraphics[width=\textwidth]{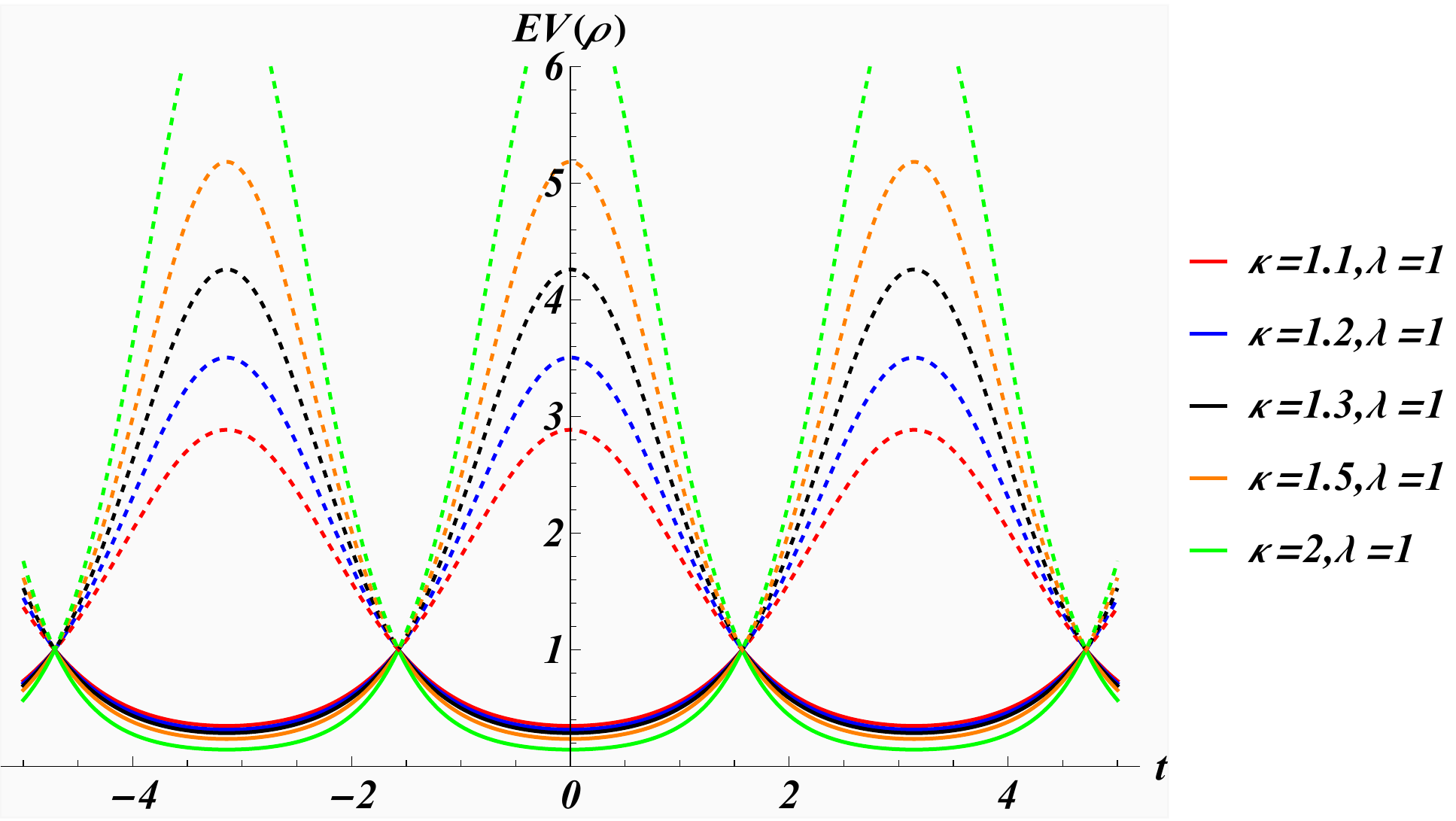}    
	\end{minipage}
	\caption{Positive-definiteness of the metric $\rho(t)$: Eigenvalues of $\rho(t)$ in (\ref{rhof}) as function of time for different values of $\kappa$ and $\lambda$ for $\tau(t) = \sin(t)$ in the ${\cal PT}$-symmetric regime, panel (a), and spontaneously broken ${\cal PT}$-regime, panel (b). Two eigenvalues corresponding to the same values of $\kappa$ and $\lambda$ are of the same colour drawn as dashed and solid lines.}
	\label{EVrho}
\end{figure}

The metric $\rho$ is indeed positive definite as its two eigenvalues are positive for all times and all values of $\kappa$, $\lambda$ in all ${\cal PT}$-regimes. Figure \ref{EVrho} displays the eigenvalues as functions of time for a specific choice of the time-dependent function $\tau(t)$ in the Hamiltonian. We observe some standard degeneracy when the metric reduces to the identity matrix at specific values of time $t=\pi /2 + n \pi$ in all regime, and in the ${\cal PT}$-symmetric regimes also at $t= \arccos( 2 n \pi / \sqrt{\lambda^2 - \kappa^2})$  with $n \in \mathbb{Z}$.

\section{Conclusions}
We have shown that time-dependent ${\cal C}$-operators can be defined in terms of solutions of the left and right Schr\"odinger equation (\ref{defCt}). However, as these solutions are related to the left and right eigenstates of the non-Hermitian Lewis-Riesenfeld invariants simply by  phase factors, they can also be expanded in terms of the latter as the phase factors of the left and right solutions cancel each other out. The three key properties of time-dependent ${\cal C}$-operators (\ref{algC}) that serve as the set up for an algebraic approach to find ${\cal C}$, have to be replaced by (\ref{algCt}) for their time-dependent versions. Since the last equation in (\ref{algCt}) is the Lewis-Riesenfeld equation it implies that time-dependent ${\cal C}$-operators are always identical to Lewis-Riesenfeld invariants. The inverse only holds when the eigenvalues of the Lewis-Riesenfeld invariants are identical to the signature of the ${\cal C}$-operators. Thus, there are plenty of Lewis-Riesenfeld invariants $I_H(t)$ that are not ${\cal C}(t)$-operators and only when we also impose the first two equations in (\ref{algCt}) does the equality $I_H(t) = {\cal C}(t)$ hold.    

Once more we have seen that in a time-dependent setting the spontaneously broken regime is mended \cite{AndTom3}, as one can find positive definite metric operators in all three ${\cal PT}$-regimes. While some of the solutions only hold in a particular regime and break down at their boundaries (\ref{rho2}), there exist also solutions (\ref{rhof}) that can be continued smoothly across all three ${\cal PT}$-regimes. 

\medskip
\noindent \textbf{Acknowledgments:} TT is supported by EPSRC grant EP/W522351/1. RT is supported by a City, University of London
Research Fellowship.

\newif\ifabfull\abfulltrue

%%\bibliographystyle{phreport}
%%\bibliography{acompat,Ref}

\end{document}